\documentclass[preprint,prl,twocolumn,10pt,showkeys,showpacs]{revtex4}%
\usepackage{aas_macros}
\usepackage{amsfonts}
\usepackage{graphicx}
\usepackage{amsmath}
\usepackage{amssymb}%
\providecommand{\U}[1]{\protect\rule{.1in}{.1in}}
\begin{document}
\title{Pair plasma relaxation time scales}
\author{A.G. Aksenov}
\affiliation{Institute for Computer-Aided Design, Russian Academy of Sciences, Vtoraya
Brestskaya 19/18, 123056 Moscow, Russia}
\author{R. Ruffini}
\affiliation{ICRANet Piazza della Repubblica, 10, 65100 Pescara, Italy}
\affiliation{ICRA and University of Rome \textquotedblleft Sapienza\textquotedblright,
Physics Department, Piazzale A. Moro 5, 00185 Rome, Italy}
\author{G.V. Vereshchagin}
\affiliation{ICRANet Piazza della Repubblica, 10, 65100 Pescara, Italy}
\affiliation{ICRA and University of Rome \textquotedblleft Sapienza\textquotedblright,
Physics Department, Piazzale A. Moro 5, 00185 Rome, Italy}

\begin{abstract}
By numerically solving
the relativistic Boltzmann equations,
we compute the time scale for relaxation to thermal equilibrium for an optically
thick electron-positron plasma with baryon loading. We focus on the time scales of
electromagnetic interactions. The collisional integrals are obtained directly from the corresponding QED matrix elements.
Thermalization time scales are computed for a wide range of values of both the total energy
density (over 10 orders of magnitude) and of the baryonic loading parameter (over 6 orders of
magnitude). This also allows us to study such interesting limiting cases as
the almost purely electron-positron plasma or electron-proton plasma as well as
intermediate cases. These results appear to be important both for laboratory
experiments aimed at generating optically thick pair plasmas as well as for
astrophysical models in which electron-positron pair plasmas play a relevant role.

\end{abstract}
\maketitle

Current interest in electron-positron plasmas is due to the exciting
possibility of generating such plasmas in laboratory facilities already
operating  or under construction, see e.g.,
\cite{2009PhRvE..79f6409M,2009RvMP...81..959T}, for a review see \cite{Ruffini2009}. Impressive progress made with ultra-intense lasers \cite{2009PhRvL.102j5001C} has led
to the creation of positrons at an
unprecedented density of $10^{16}$ cm$^{-3}$ using
ultra-intense short laser pulses, in a region of space with dimensions on the order of the Debye length.
However, such densities have not yet reached those necessary for the creation of an optically thick pair plasma \cite{2000ApJS..127..371K,2009PhRvA..79b0103M}.
Particle pairs are created at the focal point of ultra-intense lasers 
via the Bethe-Heitler
conversion of hard x-ray bremsstrahlung photons \cite{2009PhRvE..79f6409M} in the
collisionless regime \cite{1992PhRvL..69.1383W}. The approach to an optically
thick phase may well be envisaged in the near future.

Electron-positron plasmas are known to be present in compact astrophysical
objects, leaving their characteristic imprint in the observed radiation spectra
\cite{2005MNRAS.357.1377C}. Optically thick electron-positron plasmas do
indeed play a crucial role in the gamma-ray burst phenomenon
\cite{Ruffini2009,2009AIPC.1132..199R}.

From the theoretical point of view electron-positron pair plasmas are interesting because of the mass symmetry
between the plasma components. This symmetry results in the absence of both acoustic modes and
Faraday rotation; waves and instabilities in such plasmas differ significantly from asymmetric
electron-ion plasmas, see e.g. \cite{1995PhRvE..51.6079Z}. Besides, theoretical progress in understanding quark-gluon plasma in the high-temperature limit is linked to understanding QED plasma since the
results in these two cases differ only by trivial factors containing the QCD degrees of freedom (color and
flavor) \cite{2009RvMP...81..959T}.

Most theoretical considerations so far have assumed that an electron-positron plasma
is formed either in thermal equilibrium (common temperature, zero chemical
potentials) or in chemical equilibrium (nonzero chemical potentials), see 
e.g.~\cite{2009RvMP...81..959T} and references therein. However, it is necessary to
establish the time scale for actually reaching such a configuration. The only way
for particles to thermalize, i.e., reach equilibrium distributions
(Bose-Einstein or Fermi-Dirac) is via collisions. Collisions become relevant
when the mean free path of the particles becomes smaller than the spatial
dimensions of the plasma, and so the optical thickness condition is crucial for thermalization
to occur.

Thermalization (chemical equilibration) time scales for optically thick plasmas are estimated in the literature by order of magnitude arguments using essentially just the reaction rates of the dominant particle interaction processes, see 
e.g.~\cite{1981PhFl...24..102G,1983MNRAS.202..467S}.
They have been computed using various approximations. In particular, electrons have been considered ultrarelativistic, and Coulomb logarithm has been replaced by a constant.
The accurate determination of such
time scales as presented here is instead accomplished by solving the relativistic Boltzmann equations
including the collisional integrals representing all possible particle
interactions. In this case the Boltzmann equations become highly nonlinear coupled
partial integro-differential equations which can only be solved numerically.

We developed a relativistic kinetic code treating the plasma as homogeneous and
isotropic and have previously determined the thermalization time scales for an electron-positron plasma
for selected initial conditions \cite{2007PhRvL..99l5003A}. This approach was
generalized to include protons in \cite{2009PhRvD..79d3008A}. We focus only on the
electromagnetic interactions, which have a time scale of less than
$10^{-9}$ sec for our system, and therefore on the proton and leptonic
component of the plasma. The presence of neutrons and their possible equilibrium due to weak
interactions will occur only on much longer time scales.

In this paper we report on the systematic results obtained by exploring the
large parameter space characterizing pair plasmas with baryonic loading. The two
basic parameters are the total energy density $\rho$ and the baryonic loading
parameter
\begin{equation}
B\equiv\frac{\rho_{b}}{\rho_{e,\gamma}}\simeq\frac{n_{p}m_{p}c^{2}}%
{\rho_{e,\gamma}}, \label{Bpar}%
\end{equation}
where $\rho_{b}$ and $\rho_{e,\gamma}$ are respectively the total energy densities of baryons
and electron-positron-photon plasma, $n_{p}$ and $m_{p}$ are the 
proton number density and proton mass, and $c$ is the speed of light. We choose
the following range of plasma parameters
\begin{align}
10^{23}  &  \leq\rho\leq10^{33}\;\mbox{erg/cm$^3$},\label{rhorange}\\
10^{-3}  &  \leq B\leq10^{3}, \label{Brange}%
\end{align}
allowing us to also treat the limiting cases of almost pure electron-positron
plasma with $B\ll1$, and almost pure electron-ion plasma with $B\simeq
m_{p}/m_{e}$, respectively. The temperatures in thermal equilibrium
corresponding to (\ref{rhorange}) are $0.1\lesssim k_{\mathrm{B}}T\lesssim10$ MeV.

Given the smallness of the plasma parameter $g=(n_{e}\lambda_{\mathrm{D}}^{3})^{-1}%
\ll1$, where $\lambda_{\mathrm{D}}$ is the Debye length and $n_{e}$ is the electron
number density, it is sufficient to use one-particle distribution
functions. In fact, for the pure electron-positron plasma, the inequality $3\,10^{-3}\leq g\leq 10^{-2}$ holds in the region of the temperatures of interest. In a homogeneous and isotropic plasma the distribution functions $f(\epsilon,t)$ depend
on the energy $\epsilon$ of the particle and on the time $t$. We treat the plasma as
nondegenerate, neglecting neutrino channels as well as the creation and
annihilation of baryons and the weak interactions \cite{2009PhRvD..79d3008A}.

The relativistic Boltzmann equations \cite{1956DAN...107...807B,
1984oup..book.....M} for photons, electrons, positrons, and protons in
our case are
\begin{equation}
\frac{1}{c}\frac{\partial f_{i}}{\partial t}=\sum_{q}(\eta_{i}^{q}-\chi
_{i}^{q}f_{i}),\label{Boltzmann}%
\end{equation}
where the index $i$ denotes the type of particle and $\eta_{i}^{q}$, $\chi
_{i}^{q}$ are the emission and the absorption coefficients for the production
of the $i$th-particle via the reaction labeled by $q$. We account for all relevant
binary and triple interactions between electrons, positrons, photons, and
protons as summarized in Tables~\ref{pairreactions} and \ref{preactions}%
.%
\begin{table}[tbp] \centering
\begin{tabular}
[c]{|c|c|}\hline\hline
Binary interactions & Radiative and\\
& pair producing variants\\\hline\hline
{M{\o }ller and Bhabha} & {Bremsstrahlung}\\
{$e_{1}^{\pm}{e_{2}^{\pm}\longrightarrow e_{1}^{\pm}}^{\prime}$}${e_{2}^{\pm}%
}^{\prime}$ & {$e_{1}^{\pm}e_{2}^{\pm}{\leftrightarrow}e_{1}^{\pm\prime}%
e_{2}^{\pm\prime}\gamma$}\\
{$e^{\pm}{e^{\mp}\longrightarrow e^{\pm\prime}}$}${e^{\mp\prime}}$ & {$e^{\pm
}e^{\mp}{\leftrightarrow}e^{\pm\prime}e{^{\mp\prime}}\gamma$}\\\hline
Single {Compton} & {Double Compton}\\
{\ $e^{\pm}\gamma{\longrightarrow}e^{\pm}\gamma^{\prime}$} & {$e^{\pm}%
\gamma{\leftrightarrow}e^{\pm\prime}\gamma^{\prime}\gamma^{\prime\prime}$%
}\\\hline
{Pair production} & Radiative pair production\\
and annihilation & and 3-photon annihilation\\
{$\gamma\gamma^{\prime}{\leftrightarrow}e^{\pm}e^{\mp}$} & $\gamma
\gamma^{\prime}${${\leftrightarrow}e^{\pm}e^{\mp}$}$\gamma^{\prime\prime}$\\
& {$e^{\pm}e^{\mp}{\leftrightarrow}\gamma\gamma^{\prime}$}$\gamma
^{\prime\prime}$\\\hline
& $e^{\pm}\gamma${${\leftrightarrow}e^{\pm\prime}{e^{\mp}}e^{\pm\prime\prime}%
$}\\\hline\hline
\end{tabular}
\caption{Microphysical processes in the pair plasma.}\label{pairreactions}%
\end{table}%
\begin{table}[tbp] \centering
\begin{tabular}
[c]{|c|c|}\hline\hline
Binary interactions & Radiative and\\
(Coulomb scattering) & pair producing variants\\\hline\hline
$p${$_{1}{p_{2}\longrightarrow p}_{1}^{\prime}{p}_{2}^{\prime}$} &
$p${${e^{\pm}\leftrightarrow p}^{\prime}e^{\pm\prime}$}$\gamma$\\\hline
$p${${e^{\pm}\longrightarrow p}^{\prime}e^{\pm\prime}$} & $p${$\gamma
{\leftrightarrow}p^{\prime}$}$e^{\pm}e^{\mp}$\\\hline\hline
\end{tabular}
\caption{Microphysical processes in the pair plasma involving protons. For details see also \cite{Ruffini2009}.}\label{preactions}%
\end{table}%
\begin{figure}[th]
\centering
\includegraphics[width=3.5in]{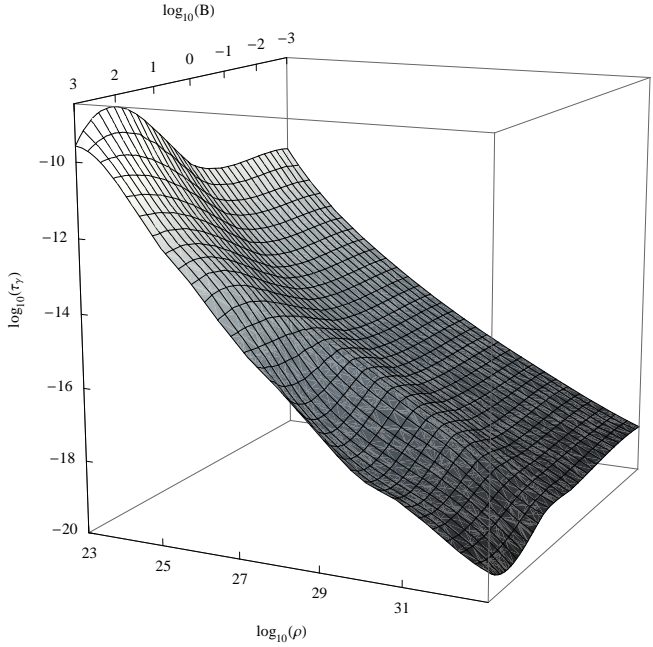} \caption{The thermalization time scale of the 
electron-positron-photon component of plasma as a function of the total energy
density and the baryonic loading parameter. The energy density is measured in
erg/cm$^{3}$, time is seconds.}%
\label{taug}%
\end{figure}\begin{figure}[th]
\centering
\includegraphics[width=3.5in]{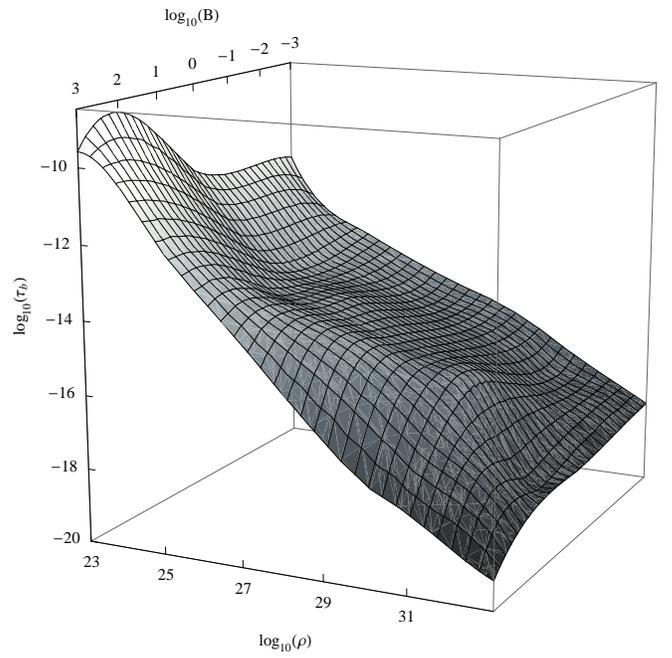} \caption{The final thermalization time scale
of a pair plasma with baryonic loading as a function of the total energy density and the
baryonic loading parameter. The energy density is measured in erg/cm$^{3}$, time
is seconds.}%
\label{taub}%
\end{figure}\begin{figure}[th]
\centering
\includegraphics[width=3in]{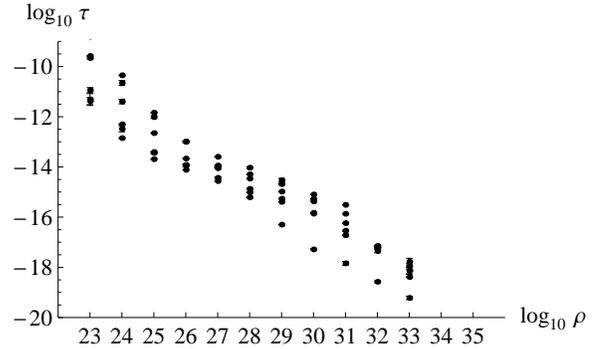} \caption{The final thermalization time scale of
pair plasma with baryonic loading as a function of the total energy density for
selected values of the baryonic loading parameter $B=(10^{-3},10^{-1.5}%
,1,10,10^{2},10^{3})$. The energy density is measured in erg/cm$^{3}$, time is
seconds. Error bars correspond to one standard deviation of the time scale
(\ref{avreltime}) away from the average value $\tau_{th}$ over the interval $t_{in}\leq t\leq t_{fin}$.}
\label{pb}%
\end{figure}

It has been shown \cite{2007PhRvL..99l5003A} that independent of the
functional form of the initial distribution functions $f_{i}(\epsilon,0)$, plasma
evolves to a thermal equilibrium state through the kinetic equilibrium, when the
distribution functions of all the particles acquire the same form
\begin{equation}
f_{i}(\varepsilon)=\exp\left(  -\frac{\varepsilon-\varphi_{i}}{\theta_{i}%
}\right)  ,\label{DFFit}%
\end{equation}
where $\varepsilon_{i}=\epsilon_{i}/(m_{i}c^{2})$ is the energy of the
particles, $\varphi_{i}\equiv\mu_{i}/(m_{i}c^{2})$ and $\theta_{i}\equiv
k_{\mathrm{B}}T_{i}/(m_{i}c^{2})$ are their chemical potentials and
temperatures, and $k_{\mathrm{B}}$ is Boltzmann's constant. The unique signature
of kinetic equilibrium is the equal temperatures of all the particles and the
nonzero chemical potential of the photons. In fact the same is also true for a pair
plasma with proton loading \cite{2009PhRvD..79d3008A}. The approach to complete
thermal equilibrium is more complicated in this latter case and depends on the
baryon loading. For $B\ll\sqrt{m_{p}/m_{e}}$, protons are rare and thermalize
via proton-electron (positron) elastic scattering, while in the opposite case
$B\gg\sqrt{m_{p}/m_{e}}$, proton-proton Coulomb scattering dominates over the
proton-electron scattering and brings protons into thermal equilibrium first with
themselves. Then protons thermalize with the pair plasma through triple
interactions, for details see \cite{2009PhRvD..79d3008A}. The two-body time scales involving protons should be compared with the
three-body time scales bringing the electron-positron-photon plasma into thermal
equilibrium. In fact we found that for $B\ll1$, the electron-positron-photon plasma
reaches thermal equilibrium at a given temperature, while protons reach
thermal equilibrium with themselves at a different temperature; only later the
plasma evolves to complete thermal equilibrium with the single temperature on
a time scale
\begin{equation}
\tau_{th}\simeq\mathrm{Max}\left[  \tau_{3p},\mathrm{Min}\left(  \tau
_{ep},\tau_{pp}\right)  \right]  ,\label{tth}%
\end{equation}
where%
\begin{align}
\tau_{ep} &  \simeq\frac{m_{p}c}{\epsilon_{e}\sigma_{\mathrm{T}}n_{e}%
},\label{tep}\\
\tau_{pp} &  \simeq\sqrt{\frac{m_{p}}{m_{e}}}\left(  \sigma_{\mathrm{T}}%
n_{p}c\right)  ^{-1},\label{tpp}\\
\tau_{3p} &  \simeq\left(  \alpha\sigma_{\mathrm{T}}n_{e}c\right)
^{-1}\label{t3p}%
\end{align}
are the proton-electron (positron) elastic scattering time scale, the proton-proton
elastic scattering time scale, and the three-particle interaction time scale
respectively, while $\sigma_{\mathrm{T}}$ is the Thomson cross-section and $\alpha$\ is
the fine structure constant. In (\ref{tep})--(\ref{t3p}) the energy dependence of
the corresponding time scales is neglected.

The chemical relaxation (thermalization) time scale is usually computed as
\begin{equation}
\tau_{i}=\lim_{t\rightarrow\infty}\left\{  \left[  F_{i}(t)-F_{i}%
(\infty)\right]  \left(  \frac{dF_{i}}{dt}\right)  ^{-1}\right\}  ,
\label{time scale}%
\end{equation}
where $F_{i}=\exp\left(  \varphi_{i}/\theta_{i}\right)  $ is the fugacity of a
particle of type $i$. Instead of $F_{i}$ we use one of the quantities
$\theta_{i}$, $\varphi_{i}$, $n_{i}$, or $\rho_{i}$ in this computation.

We solved the Boltzmann equations with parameters $(\rho,B)$ in the range
given by Eqs.~(\ref{rhorange}) and (\ref{Brange}). In total 78 models were
computed, starting from a nonequilibrium configuration until reaching a
steady state solution on the computational grid with 20 intervals for the particle
energy and 16 intervals for the angles, for details see \cite{2009PhRvD..79d3008A}.
For each model we computed the corresponding time scales for all particles of the
$i$th kind. For practical purposes, instead of (\ref{time scale}) we used the
following approximation
\begin{equation}
\tau_{th}=\frac{1}{t_{fin}-t_{in}}\int_{t_{in}}^{t_{fin}}\left[
\theta(t)-\theta(t_{\max})\right]  \left(  \frac{d\theta}{dt}\right)  ^{-1}dt,
\label{avreltime}%
\end{equation}
with $t_{in}<t_{fin}<t_{\max}$, where $t_{\max}$ is the moment of time where the steady
solution is reached and $t_{in}$ and $t_{fin}$ are the boundaries of the time
interval over which the averaging is performed, for details see
\cite{2009AIPC..........A}. 
\begin{figure}[th]
\centering
\includegraphics[width=3in]{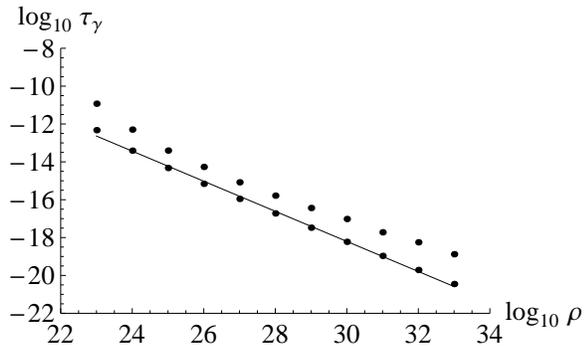} \caption{The thermalization time scale of the
electron-positron-photon component of the plasma as a function of the total energy
density (points), compared with the $\tau_{3p}$ time scale (joined points)
computed using (\ref{t3p}) for $B=1$. The energy density is measured in erg/cm$^{3}$,
time is seconds.}%
\label{ptg}%
\end{figure}\begin{figure}[th]
\centering
\includegraphics[width=3in]{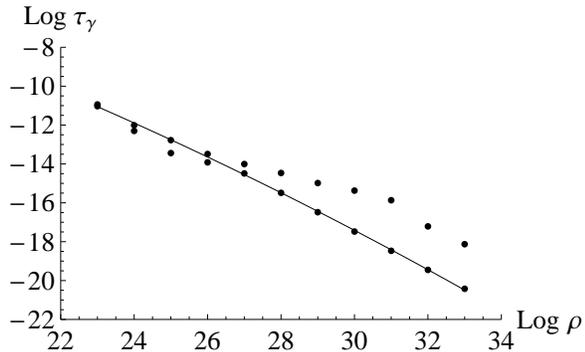} \caption{The final thermalization time scale
of a pair plasma with baryonic loading as a function of the total energy density
(points), compared with the $\tau_{th}$ time scale (joined points) computed using
(\ref{tth}) for $B=1$. The energy density is measured in erg/cm$^{3}$, time is
seconds.}%
\label{ptb}%
\end{figure}
The thermalization time scale of the electron-positron-photon component is
shown in Fig.~\ref{taug} as a function of the total energy density of the plasma and the
baryonic loading parameter. The time scales of electrons, positrons and photons
coincide. The final thermalization time scale of pair plasma with baryonic loading
is shown in Fig.~\ref{taub}. Its dependence on either variable cannot be fit by a simple
power law, although it decreases monotonically with increasing total energy
density, while it is not even a monotonic function of the baryonic
loading parameter.

In Fig.~\ref{pb} the final thermalization time scale is shown for all the models we
computed, along with the ``error bars" which mark one standard deviation of the
time scale (\ref{avreltime}) away from the average value $\tau_{th}$ in the averaging interval $t_{in}\leq t\leq t_{fin}$. The largest source of error
comes from the small values of the time derivative in (\ref{avreltime}), although
errors are typically below a few percent.

In Fig.~\ref{ptg} we compare for $B=1$ the actual value of the thermalization
time scale of the electron-positron-photon component with the value estimated from
(\ref{t3p}). Both values clearly differ significantly. Actually the
systematic underestimation by more than one order of magnitude which occurs for
$B\leq1$ disappears for larger baryonic loading.

In Fig.~\ref{ptb} we present the computed values of the final thermalization
time scale of the pair plasma with baryonic loading together with the value
estimated from (\ref{tth}), again for $B=1$. Unlike the previous case, the
final thermalization time scale is a more complex function of the total energy
density. Interestingly, less significant deviations from the value (\ref{tth})
occur at the extremes of the interval (\ref{Brange}).

In this paper we have computed for the first time the time scale of thermalization
for an electron-positron plasma with proton loading over wide ranges of both the
total energy density (10 orders of magnitude) and baryonic loading parameter
(6 orders of magnitude) allowing the treatment of the limiting cases of almost pure
electron-positron plasma, almost pure electron-ion plasma as well as
intermediate cases. The final result is presented in Fig.~\ref{taug}\ and
\ref{taub}. The relaxation to thermal equilibrium for the total energy density
(\ref{rhorange}) always occurs  on a time scale less than $10^{-9}$ sec. It is
interesting that the electron-positron-photon component and/or proton component
can thermalize earlier than the time at which complete thermal equilibrium is reached. The
relevant time scales are given and compared with the order-of-magnitude
estimates. Unlike previous work there are no simplifying assumptions in
our method since collisional integrals in the Boltzmann equations are computed
directly from the corresponding QED matrix elements, e.g. from the first principles.

These results may be of relevance for the ongoing and future laboratory experiments aimed at creating
electron-positron plasmas. Current optical lasers producing pulses during $\sim 10^{-15}$ sec carrying energy $\sim 10^2$ J$=10^{9}$ erg are capable to produce positrons with the number density $10^{16}$ cm$^{-3}$ \cite{2009PhRvL.102j5001C}. There are claims that densities of the order of $10^{22}$ cm$^{-3}$ are reachable \cite{2002PhRvE..65a6405S}. These densities today are yet far from $10^{28}$ cm$^{-3}$ required for the plasma with the size $r_0\simeq \mu m$ to be optically thick \cite{2000ApJS..127..371K}. Notice, that the expansion timescale of such plasma will be $r_0/c\sim 10^{-14}$ sec, while the timescale to establish kinetic equilibrium for the number density considered is of the same order of magnitude. These arguments show that theoretical results obtained assuming thermal or kinetic equilibrium, such as in \cite{2009RvMP...81..959T}, cannot be applied to pair plasma, generated by ultraintense lasers.

However, results presented in this paper are important for understanding astrophysical systems observed today in which optically thick electron-positron plasmas are present. As specific example we recall that electron-positron pairs play the crucial rule in the dynamics of GRB sources. Considering typical energies and initial radii for GRB progenitors \cite{1999PhR...314..575P}
\begin{equation}
10^{48}\mathrm{erg}<E_{0}<10^{54}\mathrm{erg},\quad10^{7}\mathrm{cm}%
<R_{0}<10^{8}\mathrm{cm},
\end{equation}
we estimate the range for the energy density in GRB sources
\begin{equation}
10^{23}\frac{\mathrm{erg}}{\mathrm{cm}^{3}}<\rho<10^{32}%
\frac{\mathrm{erg}}{\mathrm{cm}^{3}},
\end{equation}
which coincides with (\ref{rhorange}). As for the baryonic loading of GRBs it is typically in the lower range of (\ref{rhorange}), namely \cite{2009AIPC.1132..199R}
\begin{equation}
10^{-3}<B<10^{-2}.
\end{equation}
Such high energy density leads to large number density of electron-positron pairs in the source of GRB, of the order of
\begin{equation}
10^{30}\mbox{ cm}^{-3} <n <10^{37}\mbox{ cm}^{-3},
\end{equation}
making it opaque to photons with huge optical depth of the order of
\begin{equation}
10^{13}<\tau<10^{18}.
\end{equation}
In fact, the radiative pressure of optically thick electron-positron plasma in these systems is responsible for the effect of accelerated expansion \cite{1999A&A...350..334R,2000A&A...359..855R,2007arXiv0705.2411B,2009AIPC.1132..199R}, leading to unprecedented Lorentz factors attained $\Gamma\simeq B^{-1}$, up to $10^3$, see e.g. \cite{2009Sci...323.1688A,Izzo2010}. The role of the baryon admixture in electron-positron plasma in GRBs is to transfer internal energy of pairs and photons into kinetic energy of the bulk motion thus giving origin to afterglows of GRBs \cite{1999PhR...314..575P,2009AIPC.1132..199R}. Notice that in GRBs the timescales of thermalization are much shorter than the dynamical timescales $R_0/c\sim 10^{-3}$ sec, which implies that expanding electron-positron plasma even in the presence of baryons is in thermal equilibrium during the accelerating optically thick phase \cite{2008AIPC.1000..309A}.

After completion of this work we learned about the publication of \cite{2009arXiv0911.0118K} where work similar to ours has been performed. Between this paper and our work conceptual differences should be noted which concern the attribution of thermalization to two-body M{\o}ller and Bhabha scattering, while we have pointed out explicitly that three-body interactions play an essential role. The thermalization time scales obtained by us have been computed with reference to these three-body interactions.

\vspace{0.3in}

{\bf Acknowledgements.} We thank both anonymous referees for their comments which allowed to improve remarkably the paper.

\end{document}